\documentclass[prl,twocolumn,floatfix,aps,superscriptaddress,showpacs]{revtex4-1}
\usepackage{amsmath,amssymb,amsbsy}
\usepackage{hyperref,subfigure,braket}
\usepackage[pdftex]{graphicx}
\usepackage[utf8]{inputenc}
\usepackage{color}
\renewcommand{\vec}[1]{\boldsymbol{#1}}

\newcommand{\dg}{^\dag}

\newcommand{\num}[1]{#1\dg #1}

\begin{document}
\title{Adiabatic State Preparation of Interacting Two-Level Systems}
\author{R.~T.~Brierley}
\author{C.~Creatore}
\affiliation{TCM, Cavendish Laboratory, JJ Thomson Avenue, Cambridge, CB3 0HE, UK}

\author{P.~B.~Littlewood}
\affiliation{TCM, Cavendish Laboratory, JJ Thomson Avenue, Cambridge, CB3 0HE, UK}
\affiliation{Argonne National Laboratory, Argonne IL 60439, USA}
\affiliation{James Franck Inst., University of Chicago, Chicago IL 60637, USA}

\author{P.~R.~Eastham}
\affiliation{School of Physics, Trinity College, Dublin 2, Ireland}
\begin{abstract}
  We consider performing adiabatic rapid passage (ARP) using
  frequency-swept driving pulses to excite a collection of interacting
  two-level systems. Such a model arises in a wide range of many-body
  quantum systems, such as cavity QED or quantum dots, where a
  nonlinear component couples to light. We analyze the one-dimensional
  case using the Jordan-Wigner transformation, as well as the mean
  field limit where the system is described by a
  Lipkin-Meshkov-Glick Hamiltonian.  These limits provide
  complementary insights into the behavior of many-body systems
  under ARP, suggesting our results are generally applicable. We
  demonstrate that ARP can be used for state preparation in the
  presence of interactions, and identify the dependence of the
  required pulse shapes on the interaction strength. In general
  interactions increase the pulse bandwidth required for successful
  state transfer, introducing new restrictions on the pulse forms
  required.\end{abstract} \maketitle

Precise control of quantum mechanical systems is a sought after
feature for applications in quantum information and investigations of
many-body quantum dynamics. Discrete atomic-like systems, or qubits, can be
excited by using external field pulses that induce Rabi oscillations,
with the final state determined by the intensity and duration of the
pulse. However, this method is sensitive to fluctuations in the
driving field, transition energy and other sources of disorder
\cite{malinovsky_general_2001}. An alternative approach, which is
robust against such variations, is the use of frequency-swept
(``chirped'') pulses to perform adiabatic rapid passage (ARP). In this
method, the frequency of the driving field is swept through the
transition to be excited, implementing the Landau-Zener process for
adiabatic passage
\cite{landau_zur_1932,zener_non-adiabatic_1932}. Provided the gap
induced by the applied field is large compared with the sweep rate the
process is adiabatic, and the wavefunction is transferred from the
initial ground state to the target state with high probability. The presence of an external field creating a gap contrasts with some recent analyses of many-body Landau-Zener problems \cite{cherng_entropy_2006,altland_many_2008,keeling_collapse_2008,altland_non-adiabacity_2009,itin_dynamics_2009} in which there is no external field creating a gap and non-adiabatic effects appear.

ARP is a well-established technique in nuclear magnetic resonance,
where chirped radio frequency pulses are used to manipulate nuclear
spins \cite{garwood_return_2001}. More recently, there have been a
number of investigations into using ARP with optical pulses to control
excitons in quantum
dots \cite{malinovsky_general_2001,schmidgall_population_2010,wu_population_2011,simon_robust_2011},
including the creation of entangled
states \cite{unanyan_preparation_2001,unanyan_entanglement_2002,kis_controlled_2004,creatore_quantum_2011}. This
has coincided with growing interest in producing many-body systems
with strong light-matter interactions, such as coupled photon cavities
or polaritonic systems \cite{hartmann_quantum_2008}. A protocol such
as ARP that allows robust control of the quantum state in these
systems would enable the investigation of quantum dynamics in highly
non-equilibrium regimes
\cite{eastham_quantum_2009,tomadin_signatures_2010,
  polkovnikov_nonequilibrium_2010,greiner_collapse_2002}.

In established examples of ARP the interactions are weak on the scale
of the level splittings generated by the ARP pulse, and hence the 
former can be straightforwardly neglected. The aim of this paper is to
demonstrate how ARP may be extended to strongly-interacting regimes
where this is not the case. We consider a model of interacting
two-level systems which, by comparison to the case of uncoupled
two-level systems \cite{malinovsky_general_2001}, allows the effect of
interactions to be identified. We show that ARP remains an effective
approach in the interacting case, provided the pulse bandwidth is
sufficient to span the spectrum of the collective modes generated by
the interactions. Although our model is relatively simple, our results
are relevant across a wide range of systems, including cavity QED
systems \cite{hartmann_quantum_2008}, quantum dots \cite{gywat_dynamics_2006,creatore_quantum_2011}, superconducting
qubits \cite{blais_cavity_2004,majer_coupling_2007,larson_circuit_2010} and doped
impurities in semiconductors \cite{morley_initialization_2010}.

The model we consider consists of a set of $N$ interacting two-level
systems driven by an external field (in the rotating wave
approximation):
\begin{equation}\begin{split}
  \label{eq:hamiltonian-pretrans}
  H&=\sum_i\left[\frac
    E2(\sigma^z_i+1)+(f_i(t)\sigma^+_i+\text{h.c.})\right] \\ &\qquad - \sum_{i,j}J_{ij}\sigma^+_i\sigma^-_j\text,\end{split}
\end{equation}
where $\vec{\sigma}_i$ are the Pauli matrices for the two-level system
$i$, and $\sigma_i^{\pm}=(\sigma_i^x\pm i \sigma_i^y)/2$. In this form
the two states $\sigma^z_i=\pm 1$ are understood to correspond to the
presence or absence of an excitation of the $i${th} two-level system,
e.g., of an exciton in a particular state of a particular quantum
dot. We will also refer to the collective pseudospin $\vec S =
\sum_i\vec \sigma_i/2$, whose z-component is related to the total
excitation or occupation $n=S_z+N/2$. $f_i(t)$ is the coherent
external pulse used to perform ARP, and $J_{ij}$ is the interaction
between systems $i$ and $j$. At this stage we assume that the energy
of the excitation $E>0$ is the same for all transitions, and neglect
interactions of the form $\sigma_i^z \sigma_j^z$. This model could be
realized in precisely engineered cavity \cite{hartmann_quantum_2008}
or circuit QED \cite{blais_cavity_2004,majer_coupling_2007,larson_circuit_2010}
systems. Furthermore, a less idealized model of this form can be used
to describe many realizations of interacting qubits, such as coupled
quantum dots \cite{gywat_dynamics_2006,creatore_quantum_2011}. These
systems often exhibit disorder in the energies $E$ and interaction
strengths $J_{ij}$, but the robustness of ARP means the general
understanding we obtain of the effect of interactions is applicable.

Decomposing the driving field into amplitude and frequency
$f_i(t)=g_i(t)\exp(i\int\omega(t')dt')$, and eliminating the
instantaneous frequency from the driving term using a unitary
transformation, the Hamiltonian becomes:
\begin{equation}
  \label{eq:hamiltonian}
  \begin{split}
    H=\sum_i\left[\frac{(E-\omega(t))}2(\sigma^z_i+1)+(g_i(t)\sigma^+_i+\text{h.c.})\right]\\ - \sum_{i,j}J_{ij}\sigma^+_i\sigma^-_j\text{.}
  \end{split}
\end{equation}
For the discussion in this paper, we consider a Gaussian, linearly
chirped pulse with uniform amplitude,
\begin{equation}
  \label{eq:pulseform}
  g_i(t) = g\exp(-t^2/\tau^2)\text,\quad \omega(t)=E+\alpha t\text,
\end{equation}
where $g$ parametrizes the pulse amplitude, $\tau$ is the temporal
width of the pulse and $\alpha$ is the linear chirp.  We discuss the
pulse and system parameters in terms of the dimensionless combinations
$g\tau$, $J\tau$ and $\alpha\tau^2$ ($\hbar=1$). For $\alpha=0$,
Eq.~\eqref{eq:pulseform} becomes a Rabi pulse centered at frequency
$E$, with a pulse area proportional to $g\tau$.

In the non-interacting case where $J=0$, the use of ARP to transfer the
two-level systems from the ground state $\sigma^z=-1$ to the excited
state is well understood \cite{malinovsky_general_2001}. For $g=0$,
the energies of the two levels cross when $E+\omega(t)=0$. The
presence of the field $g\neq0$ produces an  avoided crossing and the
adiabatic state smoothly varies from the initial ground state to the
excited state. When the pulse amplitude is time independent, $g(t)=g$,
the model reduces to the canonical Landau-Zener problem
\cite{landau_zur_1932,zener_non-adiabatic_1932}. The probability of
remaining in the adiabatic state (and so being transferred from the
initial ground state to the excited state) is $1-\exp(-2\pi
g^2/\alpha)$, so that the final population is always increased by
reducing the chirp $\alpha$, increasing the adiabaticity of the
process. In the case of ARP, using pulses of finite duration, $g(t)$
is no longer constant. Thus, in order for adiabatic passage to occur,
the two levels of the system must be coupled together long enough that
the character of the eigenstates changes sufficiently slowly. This
introduces the requirement that $\alpha\gg1/\tau^2$
\cite{malinovsky_general_2001}. In the limit $\alpha\to0$, the system
undergoes Rabi oscillations rather than ARP.

In order to understand how this process generalizes to the interacting
case, we first examine a one-dimensional chain with nearest neighbor
interaction $J_{ij}=J\delta_{i,i+1}$. In this case, the energy levels
for $g=0$ can be determined using the Jordan-Wigner transformation
$\sigma^z_i=2\num{c_i}-1$,
$\sigma^-_i=\exp(i\pi\sum_{j<i}\num{c_j})c_i=T_i c_i$ where $c_i$ are
fermionic operators \cite{mahan_many-particle_1990}. After also
performing a Fourier transformation the Hamiltonian,
Eq.~\eqref{eq:hamiltonian}, becomes:
\begin{equation}
\label{eq:diag-hamil}
  H=-\sum_{k}[\alpha t+J\cos k] \num{c_k}+\frac{1}{\sqrt N}\sum_{k,i} (g^*_iT_ic_ke^{ikr_i}+\text{h.c.})\text{,}
\end{equation}
where $N$ is the number of sites and $k=-\pi+2\pi m/N$ with $m$
integer.  The Jordan-Wigner transformation has previously been used to
describe Landau-Zener transitions for anisotropic spin chains in a
changing magnetic field
\cite{cherng_entropy_2006,de_pasquale_xy_2009}. The Landau-Zener
transitions in that model result from the anisotropy, which affects
the subspaces spanned by fermion operators of a given $|k|$
independently. In contrast, the spatial dependence of the nonlinear
$T_i$ term in our model leads to terms in the Hamiltonian that couple
fermion states with different $|k|$.

\begin{figure}[htbp]
  \centering
  \includegraphics[width=8.6cm]{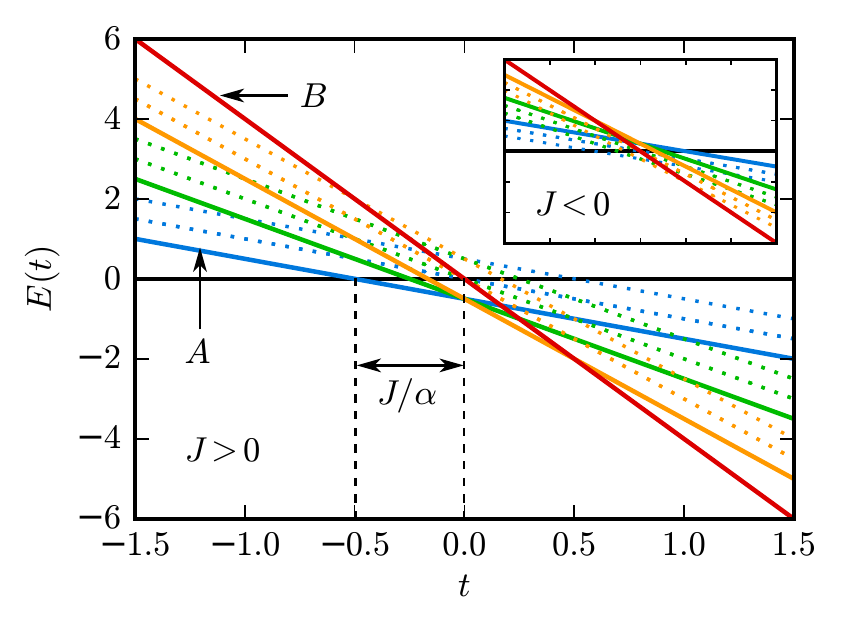}
  \caption{(Color online) Time-dependent eigenenergies of the one-dimensional chain,  Eq.~\eqref{eq:diag-hamil},
    with $g=0$, $J=\frac12$ and $N=4$ sites. Different colors
    indicate bands corresponding to different values for the occupation $n$ of the two-level systems. Bold lines
    show the states in each band with maximum $\vec S^2$, which are coupled
    together when the pump is spatially uniform.
    Dotted lines are other eigenstates of the system, which have
    different values of $\vec S^2$. The vertical dashed lines show the
    separation in time between the crossing of the $n=0$ band with the
    $n=1$ (A) and $n=N$ (B) states. Inset: As main figure but with
    $J=-\frac12$. Note that the order of crossings of the coupled
    (bold) levels has reversed. }
  \label{fig:eigenst-near-neighb}
\end{figure}

In the Jordan-Wigner representation the different energy eigenstates
correspond to different occupations of the fermion states. The
completely empty (spin down) state corresponds to the vacuum with no
fermions $\ket0$. Likewise, the completely occupied (spin up) state
corresponds to the case with all fermion states filled,
$\prod_{k}c\dg_{k}\ket0$. For large $|t|$ the first term in
Eq.~\eqref{eq:diag-hamil} dominates and the eigenstates, shown in
Fig.~\ref{fig:eigenst-near-neighb} for the few-body case $N=4$, are
split into $N+1$ bands labelled by the total number of fermions, which
physically corresponds to the excited-state population of the
two-level systems $n$. Considering the eigenstate structure from the
few-body limit is useful since, as we will show, it can be used to
understand results in the thermodynamic limit and it has been
extensively studied
\cite{brundobler_s-matrix_1993,vitanov_laser-induced_2001,ostrovsky_exact_1997,volkov_exact_2004,volkov_no-go_2005,sinitsyn_counterintuitive_2004,shytov_landau-zener_2004}.

In the non-interacting case, the energy levels are independent of $k$
and all the states in the $n$th band have energy $-n\alpha t$. The
presence of interactions, $J\neq0$, lifts the complete degeneracy of
states within each band as shown in
Fig.~\ref{fig:eigenst-near-neighb}. The separate states correspond to
the different allowed values of the total spin $\vec S^2$ for a given
$S^z$. In order to prepare a fully occupied state, the quantum state
must then be transferred via multiple level crossings from the $n=0$
to $n=N$ bands \cite{brundobler_s-matrix_1993,witthaut_towards_2006}.

The splitting of the level crossings that allows adiabatic state
transfer is introduced by the external pulse field $g$. In the general
case there will be some variation in the driving field between the
two-level systems, which makes the form of the coupling term in the
fermionic representation complicated to determine due to the
non-locality of the Jordan-Wigner string $T_i$. However, for the
uniform driving we consider [$g_i(t)=g(t)$], the coupling to the field
in the untransformed Hamiltonian, Eq.~\eqref{eq:hamiltonian}, can be
rewritten $g(t)\sum_i(\sigma^+_i+\sigma^-_i)=g(t)(S^++S^-)$, and the
transitions therefore conserve $\vec S^2$. If the system starts in the
ground state, it will thus always be in an eigenstate of $\vec S^2$
with the maximal eigenvalue $S(S+1)$ where $S=N/2$. There is one state
in each band with this value of $\vec S^2$, and the transitions
between these states have matrix elements
$g\sqrt{(N-n)\left[(n+1/2)\pm 1/2\right]}$. The corresponding states
in each band are the most symmetrical states, which for $J>0$ ($J<0$)
have the lowest (highest) energies, see
Fig.~\ref{fig:eigenst-near-neighb} (Fig.~\ref{fig:eigenst-near-neighb}
inset) \footnote{ In the opposite limit of a field applied at only one
  site, each level is coupled to all others in the adjacent bands.}. The finite $N$ model in this limit is then similar to one used to describe adiabatic control of rotational states in molecules \cite{vitanov_adiabatic_2004}.

The field term in the Hamiltonian only changes the number of fermions,
$n$, by $\pm1$. An avoided crossing between non-adjacent bands can be
induced by higher order virtual transitions. For example, a $n\to
(n+2)$th band transition is possible via an intermediate $(n+1)$th
band state. These higher-order interactions are suppressed in the $N\to\infty$ 
mean-field limit discussed below
\cite{witthaut_towards_2006,trimborn_nonlinear_2010}, but do play a
role in ARP if the connectivity is small \cite{vitanov_adiabatic_2004}. 

\begin{figure}[htbp]
  \centering
  \includegraphics[width=8.6cm]{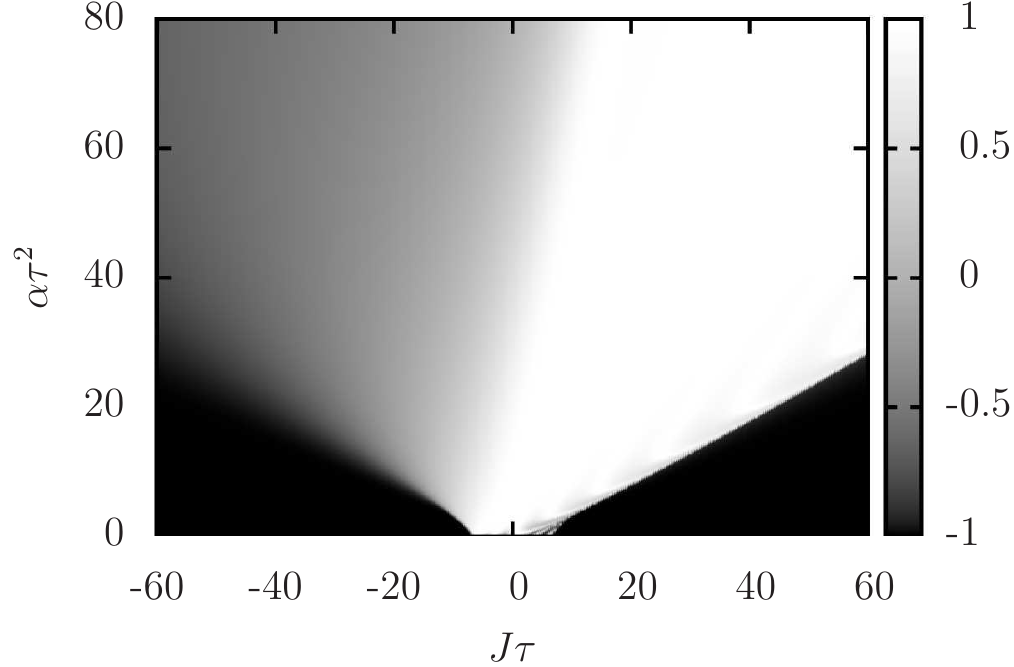}
  \caption{Mean-field calculation of the average excitation of a set
    of interacting two-level systems, Eq.~\eqref{eq:hamiltonian},
    driven from its ground state by the chirped pulse,
    Eq.~\eqref{eq:pulseform} with $g\tau=3$. $J\tau$ is the
    dimensionless interaction strength, and $\alpha\tau^2$ the
    dimensionless chirp. The regions where no excitations are created
    are a result of the finite duration of the pulse $\tau$: as the
    chirp $\alpha$ is reduced, the level crossings of
    Fig.~\ref{fig:eigenst-near-neighb} no longer occur within the
    pulse and so adiabatic transfer is not possible. The boundary of
    this region is approximately $\alpha\propto J$.}
  \label{fig:inversion-two-level}
\end{figure}

The Jordan-Wigner transformation is only usefully applicable for the
special case of nearest-neighbor hopping in one dimension. In higher
dimensions, an alternative approach is to use the mean field
approximation, which is exact in the limit $N\to\infty$,
$J_{ij}=J/N^2$. We show in Fig.~\ref{fig:inversion-two-level} results
for the final occupation obtained using a spatially uniform pulse,
Eq.~\eqref{eq:pulseform}, 
calculated by solving the Heisenberg equations of motion using the
mean field replacement
$\sum_{ij}J_{ij}\sigma^+_i\sigma^-_j=\sum_iJ_{\text{eff}}(\sigma^+_i\left<\sigma^-_i\right>+h.c.)$. In
this approximation the Hamiltonian, Eq.~\eqref{eq:hamiltonian}, can be
rewritten in terms of the collective spin operators as the
Lipkin-Meshkov-Glick Hamiltonian:
\begin{gather}
  \label{eq:mean-field-hamil}
  \begin{aligned}
  H_{MF}&=-J_{\mathrm{eff}}(S^+ S^-+S^-S^+)-\frac{\alpha t}2 S^z+2g S^x \\ 
  &=2J_{\mathrm{eff}}(S^z)^2-\frac{\alpha t}{2}S^z+2g(t)S^x,
\end{aligned}
\end{gather}
where we have used $S^2=(S^z)^2+(S^+S^-+S^-S^+)/2$  and
dropped terms which do not affect the dynamics. This Hamiltonian, with a
time independent $g(t)=g$, has been used to describe Landau-Zener tunneling for
a bosonic Josephson junction
\cite{wu_nonlinear_2000,liu_theory_2002,witthaut_towards_2006,zobay_time-dependent_2000,chen_many-body_2011,trimborn_nonlinear_2010}.

Figure~\ref{fig:inversion-two-level} consists of a fan of non-zero
occupation bounded by large regions of essentially zero occupation. The features of this result can be understood by using the intuition from considering a finite set of $N$ level crossings, as illustrated in Fig.~\ref{fig:eigenst-near-neighb}. 
Within the fan, as in the normal Landau-Zener problem, increasing
$\alpha$ decreases the final occupation as the increased velocity of
the level crossing reduces the adiabaticity of the transition.

For a fixed value of $\alpha$, the occupation within the fan increases
(decreases) for $J>0$ ($J<0$). This variation corresponds to the
changing relative positions of the level crossings, visible in
Fig.~\ref{fig:eigenst-near-neighb} for the one-dimensional chain.  In
the absence of interactions, $J=0$, the states within each band are
degenerate and so all levels cross simultaneously at $t=0$. As the
interaction strength $J$ is increased, the level crossings separate in
time. As each level degeneracy becomes more isolated, the size of the
avoided crossing caused by $g$ increases.  When $J>0$ and $\alpha>0$ the
crossings occur in ``ascending order'', i.e.~the $n\to n+1$ crossing
occurs before the $n+1\to n+2$ crossing. The increase in splitting due
to the isolation of crossings then improves the efficiency of transfer
to the occupied state, producing the increase of the occupation shown
in Fig.~\ref{fig:inversion-two-level}. If $J<0$, however, then for
positive chirp the crossings occur in the ``wrong order'' (see
Fig.~\ref{fig:eigenst-near-neighb}, inset), so that it becomes more
difficult for the system state to transfer via a series of transitions
through adjacent bands, suppressing the probability of full
occupation. In the mean field limit, this ordering leads to the
formation of a swallowtail in the energy level evolution which causes
a breakdown of adiabaticity and a corresponding reduction in the
occupation
\cite{wu_nonlinear_2000,zobay_time-dependent_2000,liu_theory_2002,witthaut_towards_2006}.

The large regions of zero occupation that define the fan are a result
of the time dependence of the optical pulse $g(t)$ used to perform
ARP. In order for the state to be transferred at a level crossing, the
avoided crossings caused by the field $g(t)$ must be large enough compared
to the level velocity $\alpha$ to make the process adiabatic. Because
ARP uses pulses of a finite duration there is only a
limited window during which $g(t)$ meets this
criterion. With no interaction, $J=0$, all crossings occur
simultaneously and so may all occur within the window. In the presence
of interactions, eventually the time of the first level crossing
$J/\alpha\sim \tau$, and it will be pushed out of the pulse.
Neglecting the change in avoided crossing size discussed above and the
effect of higher order, virtual transitions, the crossover to
non-adiabatic behavior occurs along a line $J\propto\alpha$, which is
approximately what is seen in Fig.~\ref{fig:inversion-two-level}.
Below this line, the behavior is no longer adiabatic and the system
undergoes more complicated dynamics, reducing to nonlinear Rabi
oscillations for $\alpha=0$. Thus a pulse used to perform ARP in a
large system must have a sufficient duration $\tau$ that it includes
the entire region of level crossings separated by the interaction.

In conclusion, we have shown the consequences of inter-system
interaction on using ARP to fully occupy an ensemble of many two level systems.
The interaction lifts the degeneracy of the eigenstates of the
Hamiltonian, Eq.~\eqref{eq:hamiltonian-pretrans}, and causes the level
crossings at which state transfer occurs to separate in time. As in
the bosonic Josephson junction
\cite{wu_nonlinear_2000,zobay_time-dependent_2000,liu_theory_2002,witthaut_towards_2006},
the isolation of each degeneracy increases the effective splitting so
that adiabatic transfer can be achieved for larger chirps than in the
non-interacting system. However, the separation of level crossings
also introduces the additional condition for ARP that the pulse
duration (or chirp) should be large enough to include all the
necessary crossings, increasing with $J$. Physically, this occurs
because the interactions broaden the spectrum into a set of collective
modes forming a path from the ground to final states, and the pulse
must cover this spectrum for the state preparation to be effective.

Although in this paper we have focused on the ideal case of uniform
$E$, $J$ and coupling $g_i$, our results apply more
generally. Fluctuations in $E$ and $J$ will change the energies and
character of the intermediate states so that they are not delocalized
across the system. However, for $|t|\to\infty$, the highest and lowest
states remain the empty or full states, so our results will still
apply. Variation in $g_i$ changes the size of splittings at a level
crossing but, with the exception of fine-tuned cases, avoided crossings
will still form, allowing adiabatic transfer.

As the model discussed in this paper represents limits of more
complicated many-body systems including the Bose-Hubbard, Dicke or
Jaynes-Cummings-Hubbard models
\cite{sachdev_quantum_2001,blais_cavity_2004,hartmann_quantum_2008},
these results can be used as a basis for understanding the behavior
of ARP in these models. It could then be used as a robust method of preparing
far-from-equilibrium states in those systems for use in quantum
information contexts or as equivalents of the quantum
quenches performed in ultracold atomic gases.

R.T.B. acknowledges useful discussions with J. Keeling. We acknowledge support from EPSRC GB grant EP/F040075/1 (C.C.),
Science Foundation Ireland grant 09/SIRG/I1592 (P.R.E.) and DOE~grant~FWP~70069~(P.B.L.)

\end{document}